\documentclass[a4paper]{PoS}

\usepackage{amsmath}
\usepackage{wasysym}
\usepackage{multirow}
\usepackage{slashed,bbm}
\usepackage{dsfont}

\newcommand*{\vcenteredhbox}[1]{\begingroup
\setbox0=\hbox{#1}\parbox{\wd0}{\box0}\endgroup}

\title{RG analysis at higher orders in perturbative QFTs\\
 in CMP}

\ShortTitle{RG analysis at higher orders in perturbative QFTs in CMP}

\author{\speaker{Nikolai Zerf}
          \thanks{Preprint number:~HU-EP-17/27}\\
                 Humboldt-Univerisit\"at zu Berlin\\
                 Institut f\"ur Physik\\
                 Newtonstra\ss e 15\\
                 D-12489 Berlin, Germany\\

        E-mail:        \email{zerf@physik.hu-berlin.de}}

\abstract{In this contribution we report on the perturbative determination of $\beta$-functions and anomalous dimensions for the chiral Ising, chiral XY and chiral Heisenberg Gross-Neveu-Yukawa model
around $D=4$ dimensions at four loops and the first Pad\'e extrapolation of critical exponents at non-trivial, infrared stable fixed points to $D=3$ to this order.
This talk is based on Ref.~\cite{Zerf:2017zqi}.
}

\FullConference{13th International Symposium on Radiative Corrections\\
		24-29 September, 2017\\
		St. Gilgen, Austria}

\begin{document}

\section{Introduction}
 The Gross-Neveu-Yukawa (GNY) models constitute a UV completion for the Gross-Neveu (GN) models in $D=4$ dimensions.
 The latter describe fermionic systems with a four fermion interaction and are extensively used as effective models in fermionic condensed matter systems.
 Depending on the realized global symmetry distinct models can be written down.
 In the case where the fermions couple via a single, real (for example z-) component of their spin to each other, 
 we obtain the Ising type of the GN model.
 In case one couples them via a scalar product of their three dimensional (real) spin vector, one is dealing with the Heisenberg type of the GN model.
 A single complex component coupling leads to the XY type.
 
 Within the given models one is interested in the determination of critical exponents of the coupling and fields at the infrared (IR) fixed point.
 
 It turns out that the GN models exhibit strong dynamics in $D=3$, 
 that means a perturbative expansion in a small coupling constant will not allow for a direct determination of
 critical exponents at the IR fixed point in $D=3$.
 
 In order to directly solve the given strong problems in $D=3$ one can apply Monte Carlo (MC)~\cite{Chandrasekharan:2013aya,Wang:2014cbw,Li:2014aoa,Huffman:2017swn,Hesselmann:2016tvh},
 Conformal Bootstrap (CBS)~\cite{Bashkirov:2013vya,Poland:2016chs,Iliesiu:2015akf,Iliesiu:2017nrv}, 
 (non-perturbative) Functional Renormalization Group (FRG)~\cite{Janssen:2014gea,Vacca:2015nta,Knorr:2016sfs,Gies:2017tod,Knorr:2017yze} methods or determinations within the large $N$ expansion~\cite{Gracey:1990wi,Gracey:1992cp,Derkachov:1993uw,Vasiliev:1992wr,Vasiliev:1993pi,Gracey:1993kb,Gracey:2017fzu,Manashov:2017rrx},
 where $N$ is the number of fermion copies.
 
 A more indirect way -- which we follow here -- is to apply perturbation theory in small couplings for the UV completed model of GN around $D=4$ and then continue the results for critical exponents back to $D=3$.
 This we can do under the assumption that physical observables like critical exponents do exist in $D$ dimensions.
 The agreement of predictions within the original GN model and its UV completion (the corresponding GNY model) is guaranteed by the fact that they belong to the same universality class
 and thus share the same symmetries and degrees of freedom in the relevant low energy region. 
 A similar perturbative calculation for the GN model in $D=2$ dimensions was carried out in Ref.~\cite{Gracey:2016mio}.
 
 In four dimensions the four fermion interactions contained in GN models are not renormalizeable and should be interpreted as effective interaction
 of an additional, unresolved bosonic scalar field.
 The latter is exchanged between pairs of two fermions via Yukawa type interactions.
 This interaction promotes the GN model to the GNY model
 after one adds a kinetic term for the scalar to make it dynamical and grants it at least one quartic scalar interaction.
 In condensed matter physics one says that the GNY model is obtained from GN model via bosonization. 

 Because the new scalar field has to obey the global symmetry of the fermion interaction the number and type of components is fixed.
 In the case of the Ising setup it is a single real valued field. In the case of the Heisenberg model it is a real three-vector field.
 For XY we have a single complex component field.
 
 A very special feature of the given models is that the fermions obey a linear dispersion relation without a gap (no mass).
 Such a dispersion relation can be realized in a honey comb lattices at the Dirac point.
 That means that the Lagrangian looks quasi relativistic and has Lorenzian symmetry.
 Although this symmetry is rather emergent in the low energy regime of the described solid state systems and not exact,
 it allows us to apply computational methods developed for high energy particle physics, where one assumes that Lorentz symmetry is an exact symmetry of nature.
 Because the fermions have no mass/gap and we do not include any terms in the Lagrangian with odd number of scalars (except of the Yukawa coupling).
 There is another global chiral symmetry, which will
 keep the fermions massless and ensure the absence of interactions involving any odd number (greater than one) of scalars when taking quantum corrections into account.

\section{Lagrangians for GNY}
The action for the GNY models is given by:
\begin{align}\label{eq:lag}
S_\lambda=\int d\tau d^{D-1}x \, ( \mathcal{L}_\psi + \mathcal{L}_{\psi\phi,\lambda})\,.
\end{align}
Where the index $\lambda\in\{\chi_I,\chi_{XY},\chi_{H}\}$ keeps track of the three different global symmetries leading to the chiral Ising, chiral XY and chiral Heisenberg GNY model.
All GNY models that are discussed here share the same kinetic term $\mathcal{L}_\psi$ for $N$ fermions:
\begin{align}\label{eq:Dirac}
	\mathcal{L}_\psi=\bar\psi(x)\slashed{\partial} \psi(x)\,.
\end{align}
Here we use the shorthand $\slashed{\partial}=\gamma_{\mu}\partial_{\mu}$ where the $\gamma$'s obey a four-dimensional representation of the Clifford algebra, $\{\gamma_\mu,\gamma_\nu\}=2\delta_{\mu\nu}\mathds{1}_4$, with $\mu, \nu, = 0,1,...D-1$.

The chiral Ising model has a global $Z_2$ symmetry ($\phi \in \mathbb{R}$) and the remaining terms in the action read
\begin{align}\label{eq:chiI}
	\mathcal{L}_{\psi\phi,\chi_I}=g\phi\bar\psi\psi+\tfrac{1}{2}\phi(m^2-\partial_\mu^2)\phi+\lambda\phi^4\,.
\end{align}

The chiral XY model has a global $U(1)$ symmetry ($\phi \in \mathbb{C}$)
\begin{align}\label{eq:chiXY}
	\mathcal{L}_{\psi\phi,\chi_{XY}}=g\phi\bar\psi P_+\psi+g\phi^*\bar\psi P_-\psi+|\partial_\mu \phi|^2+ m^2 |\phi|^2+\lambda|\phi|^4\,.
\end{align}
Here $P_\pm=\frac{1}{2}(1\pm\gamma_5)$ an $\gamma_5$ is kept naively anticommuting $\{\gamma_5,\gamma_\mu\}$=0.

For the chiral Heisenberg model the global symmetry is of $SU(2)$ type ($\phi \in \mathbb{R}^3$)
\begin{align}\label{eq:chiH}
	\mathcal{L}_{\psi\phi,\chi_H}=g\,\bar\psi (\vec{\phi}\cdot\vec{\sigma})\psi+\tfrac{1}{2}\vec{\phi}\cdot\left(m^2-\partial_\mu^2\right)\vec{\phi}+\lambda\left(\vec{\phi}\cdot\vec{\phi}\right)^2\,.
\end{align}
Here $\vec{\sigma}$ is the vector of three Pauli matrices which are proportional to the generators of $SU(2)$ in the fundamental representation.

For all three models one faces a spontaneous breakdown of the specific global symmetry when the mass parameter $m^2$ becomes smaller than zero,
because the scalar potential is not at its minimum value for $\phi=0$.
Choosing the state with minimal energy to be the ground or vacuum state $|0\rangle$ of the corresponding QFT then leads to a non-vanishing vacuum expectation value $\langle0|\phi|0\rangle=v$ (VEV) for the corresponding scalar field.
When one assumes that the parameter $m^2$ can be freely tuned (depending on some macroscopic parameter of the system) one thus observes a Quantum Phase Transition (QPT) whenever $m^2$ goes through the critical value zero.
In the following we are interested in the critical exponents of the fields and couplings at the Quantum Critical Point (QCP) $(m^2=0)$ of such a phase transition.

\section{Z-Factors in Perturbation Theory}
In order to perturbatively calculate quantum corrections to the critical exponents at the phase transition,
one needs to regularize and renormalize the given Lagrange densities in a first step.
For convenience we choose Dimensional Regularization (DREG) as regulator and employ the commonly used $\overline{\rm{MS}}$ scheme for the subtraction of UV divergences.

Because the given models are renormalizeable around $D=4$ all appearing UV divergences -- regulated in terms of poles in small $\epsilon$ ($=4-D$) -- can be absorbed in $Z$-factors via a redefinition of
parameters and fields:
\begin{align*}
 \phi\rightarrow \phi^0=&\sqrt{Z_{\phi}}\phi\,,& \psi\rightarrow \psi^0=&\sqrt{Z_{\psi}}\psi\,,&   &  \\
 \lambda\rightarrow \lambda^{0}=&\mu^{\epsilon}Z_{\lambda}\lambda\,, & g\rightarrow g^{0}=&\mu^{\epsilon/2}Z_{g}g\,,& m^2\rightarrow  (m^{2})^0=&Z_{m^2} m^2
\end{align*}
At tree-level all $Z-$factors are equal to one.
With growing order in perturbation theory the $Z$-factors depend polynomial on the self coupling $\lambda$ and Yukawa coupling $g$ to higher powers.
Plugging in the redefinition of fields and couplings in the Lagrangian, using the shorthands 
\begin{align*}
	Z_{\phi^4}= & Z_{\lambda} Z_{\phi}^2\,,\qquad Z_{\psi\psi\phi}= Z_{g} \sqrt{Z_{\phi}}Z_{\psi}\,,\qquad Z_{\phi^2}= Z_{\phi}Z_{m^2}\,,
\end{align*}
one obtains 

  \begin{align}
   \mathcal{L}_{\chi_I}=&Z_\psi\mathcal{L}_\psi +Z_{\psi\psi\phi}g\phi\bar\psi\psi+\tfrac{1}{2}\phi(Z_{\phi^2}m^2-Z_\phi\partial_\mu^2)\phi+Z_{\phi^4}\lambda\phi^4\,.\\
   \mathcal{L}_{\chi_{XY}}=&Z_\psi\mathcal{L}_\psi +Z_{\psi\psi\phi}g(\phi\bar\psi P_+\psi+\phi^*\bar\psi P_-\psi)\nonumber\\
   &+Z_\phi|\partial_\mu \phi|^2+ Z_{\phi^2}m^2 |\phi|^2+Z_{\phi^4}\lambda|\phi|^4\,.\label{EQ:LXY}\\
   \mathcal{L}_{\chi_H}=&Z_\psi\mathcal{L}_\psi +Z_{\psi\psi\phi}g\,\bar\psi (\vec{\phi}\cdot\vec{\sigma})\psi\nonumber\\
   &+\tfrac{1}{2}\vec{\phi}\cdot\left(Z_{\phi^2}m^2-Z_\phi\partial_\mu^2\right)\vec{\phi}+Z_{\phi^4}\lambda\left(\vec{\phi}\cdot\vec{\phi}\right)^2.
  \end{align}

\begin{figure}
\centering
\begin{tabular}{cccc}
 \vcenteredhbox{\includegraphics[scale=0.22]{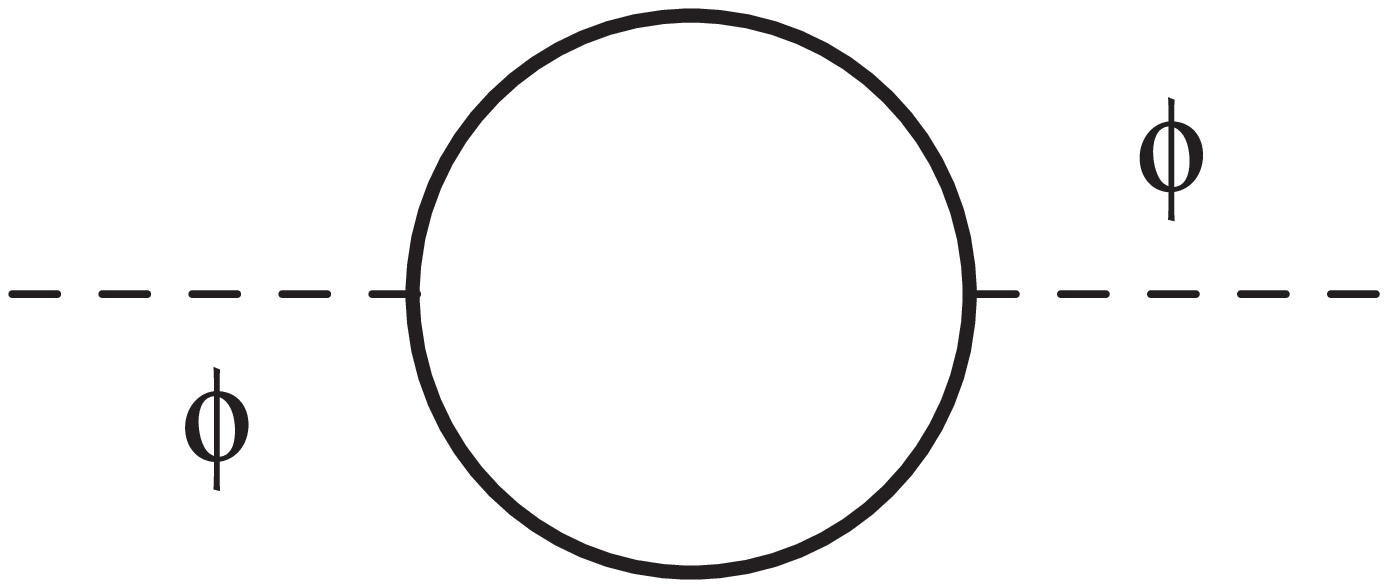}} & \vcenteredhbox{\includegraphics[scale=0.22]{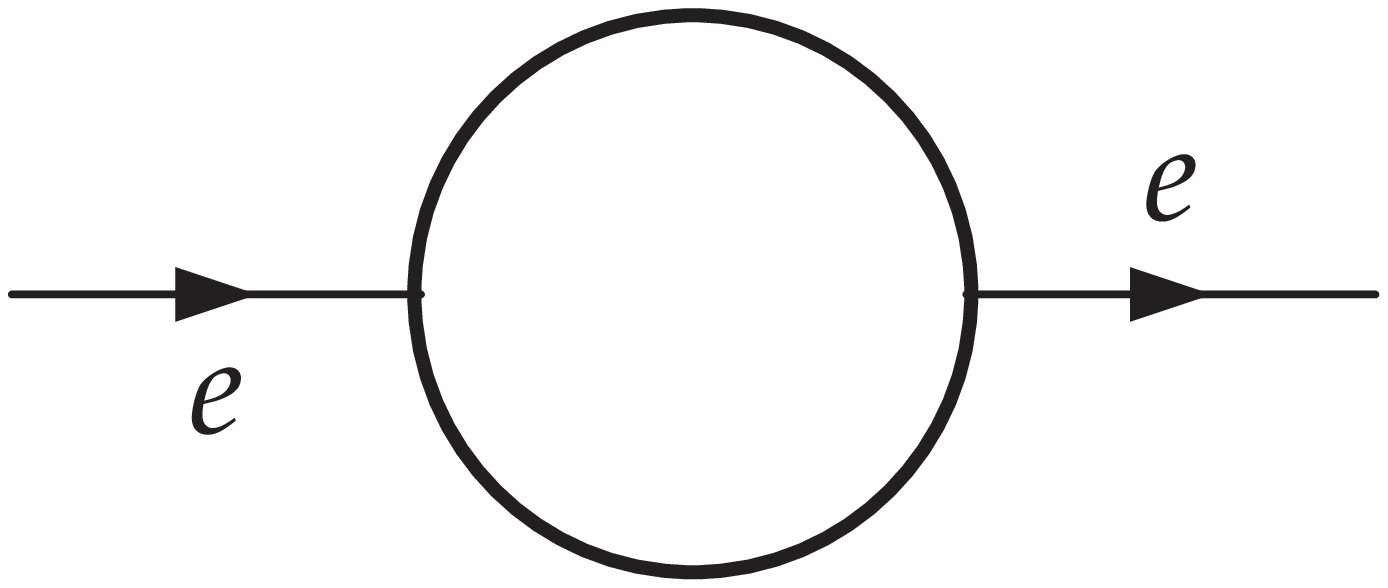}} &
 \vcenteredhbox{\includegraphics[scale=0.22]{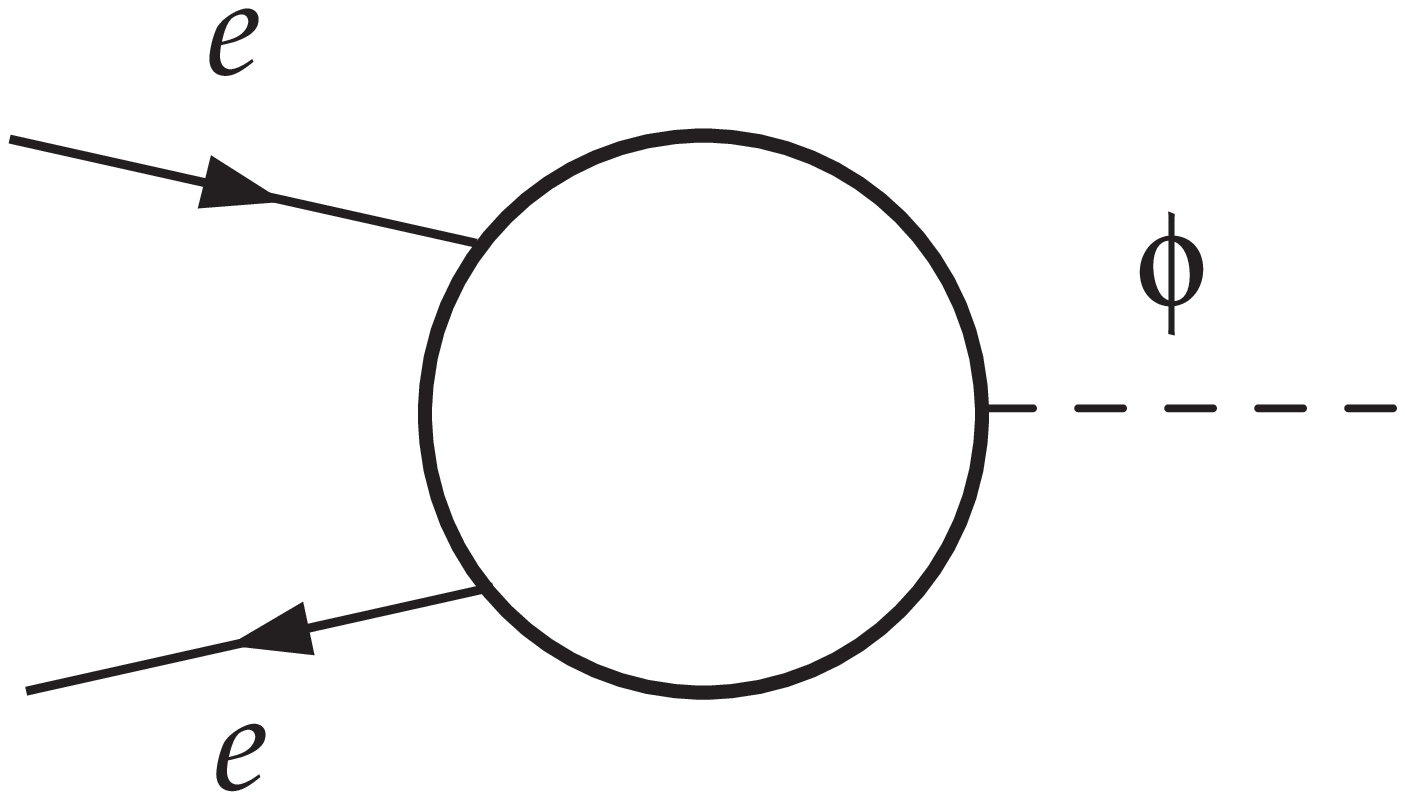}} & \vcenteredhbox{\includegraphics[scale=0.22]{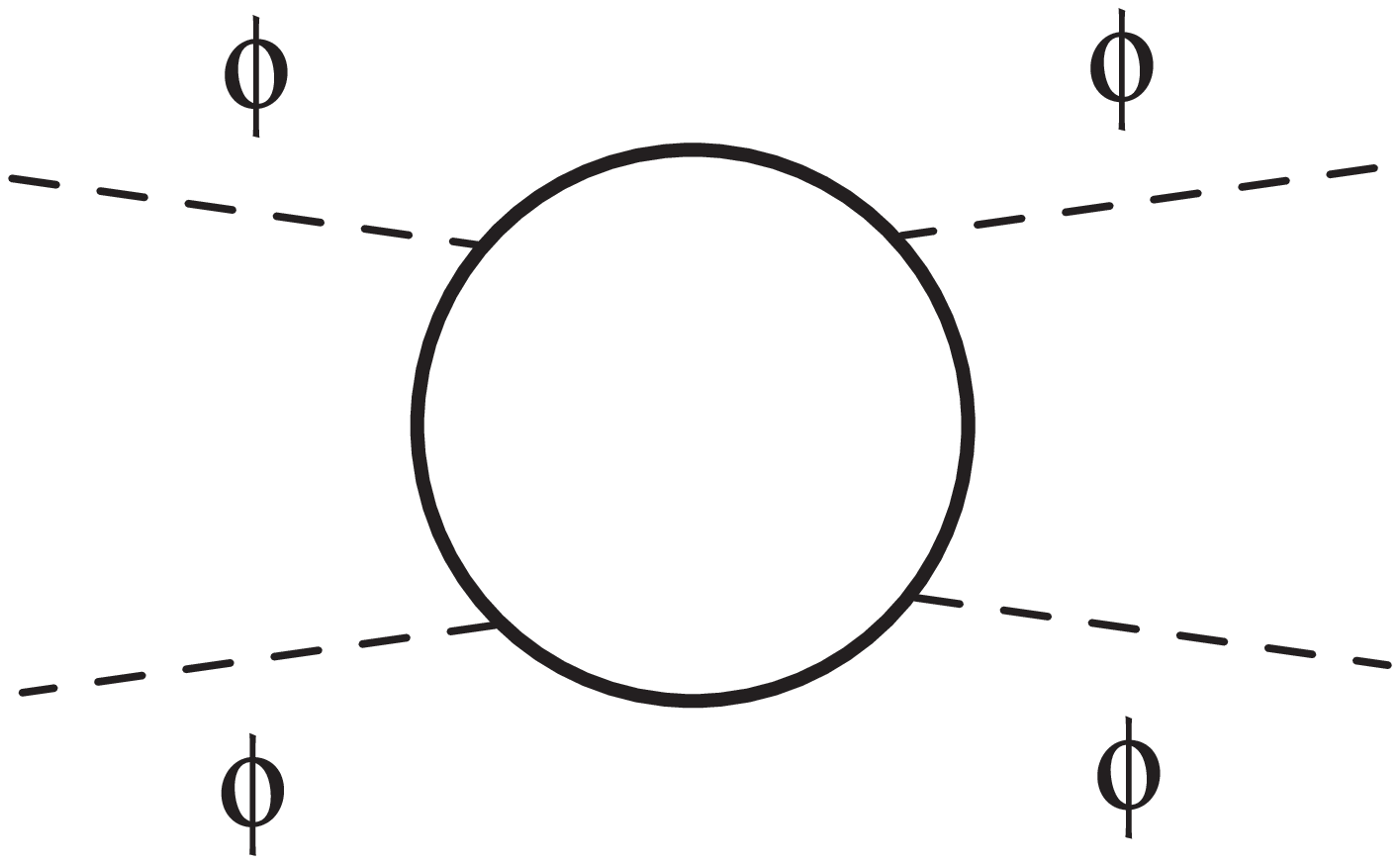}}\\
 $\sim Z_{\phi^2},Z_{\phi}$ & $\sim Z_{\psi}$ & $\sim Z_{\psi\psi\phi}$ & $\sim Z_{\phi^4}$
\end{tabular}
\caption{ Considered $n$-point Green function for the extraction of required $Z$-factors. The white blob indicates the sum of all possible 1-PI $L$-loop graphs. The $\phi$/$e$s label scalar boson/fermion lines.\label{FIG:nPgreen}}
\end{figure}

\begin{figure}
\centering
\begin{tabular}{cccc}
 \vcenteredhbox{\includegraphics[scale=0.22]{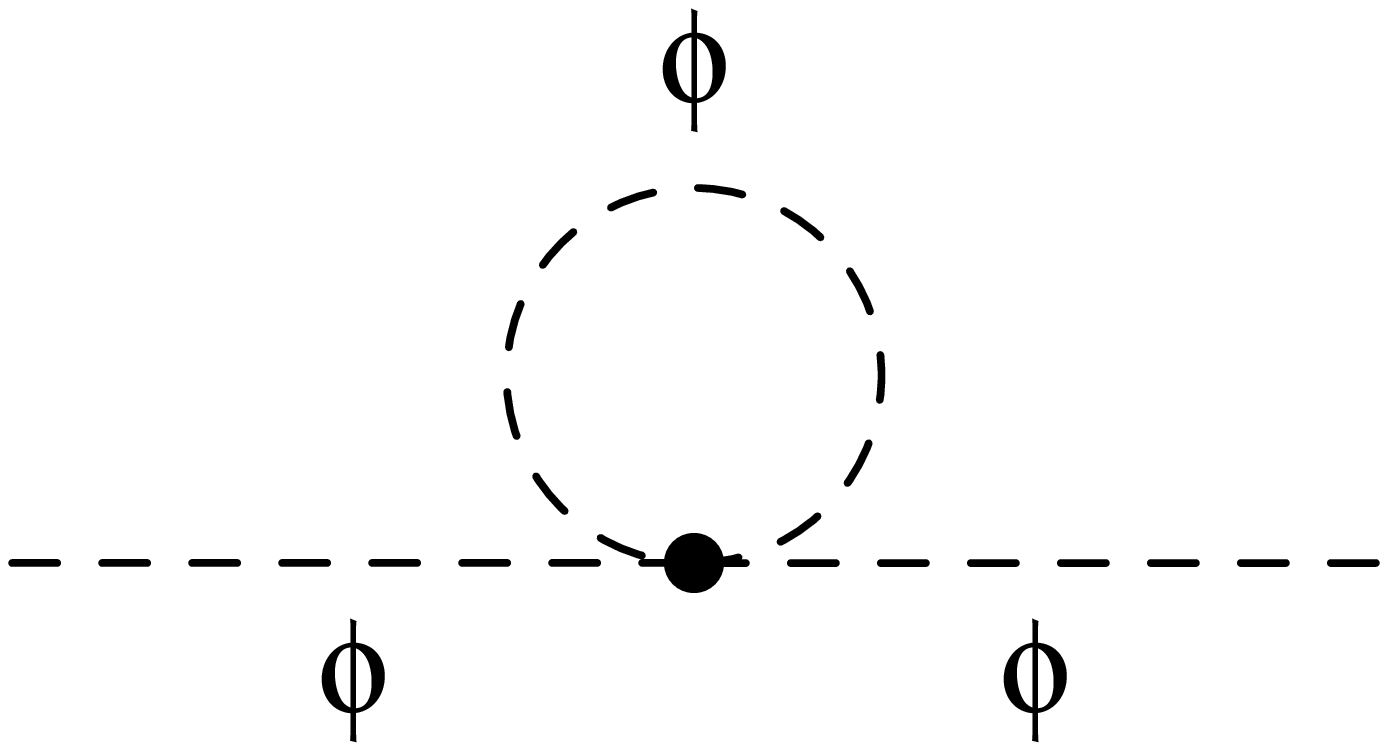}} & \vcenteredhbox{\includegraphics[scale=0.22]{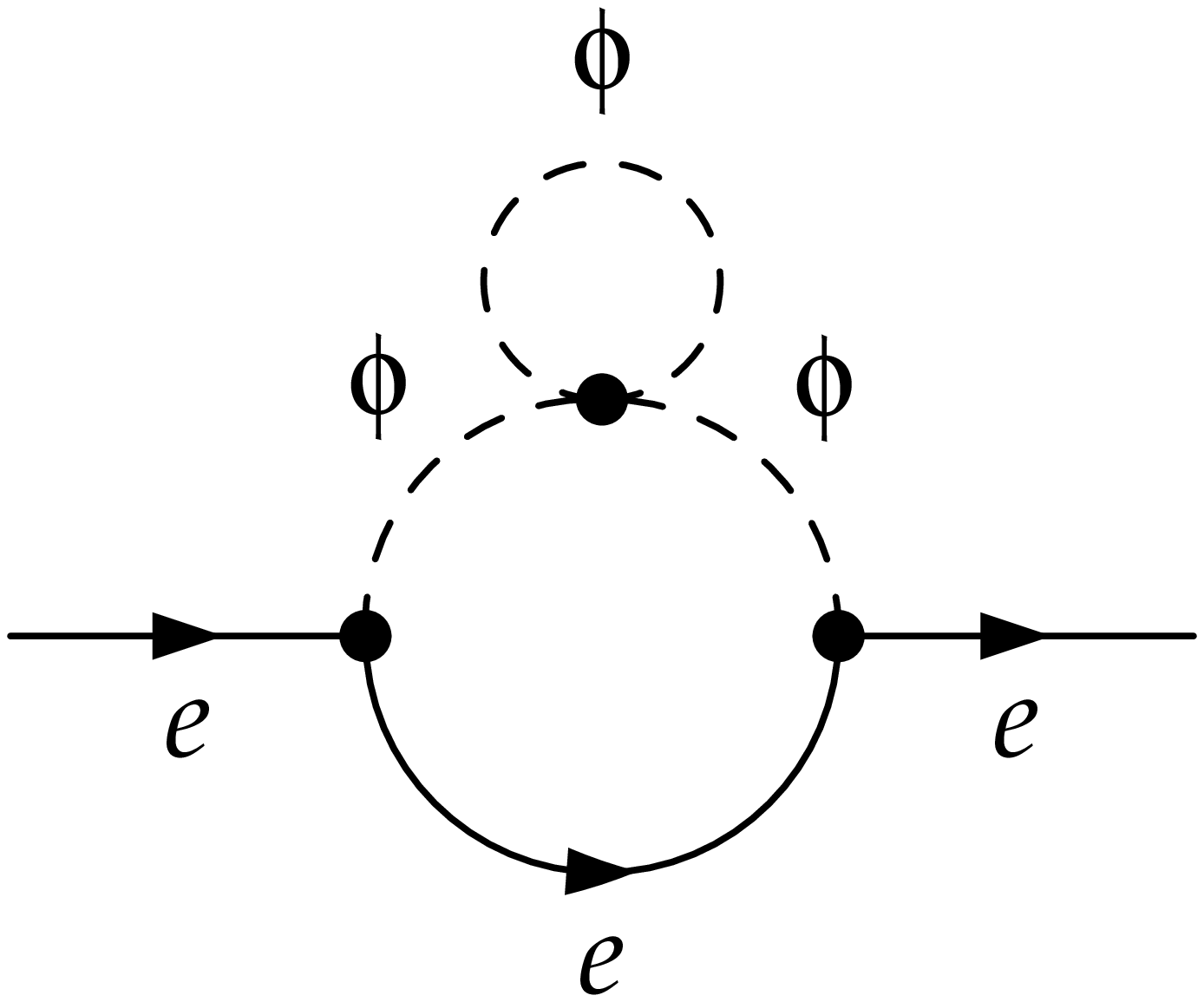}} &
 \vcenteredhbox{\includegraphics[scale=0.22]{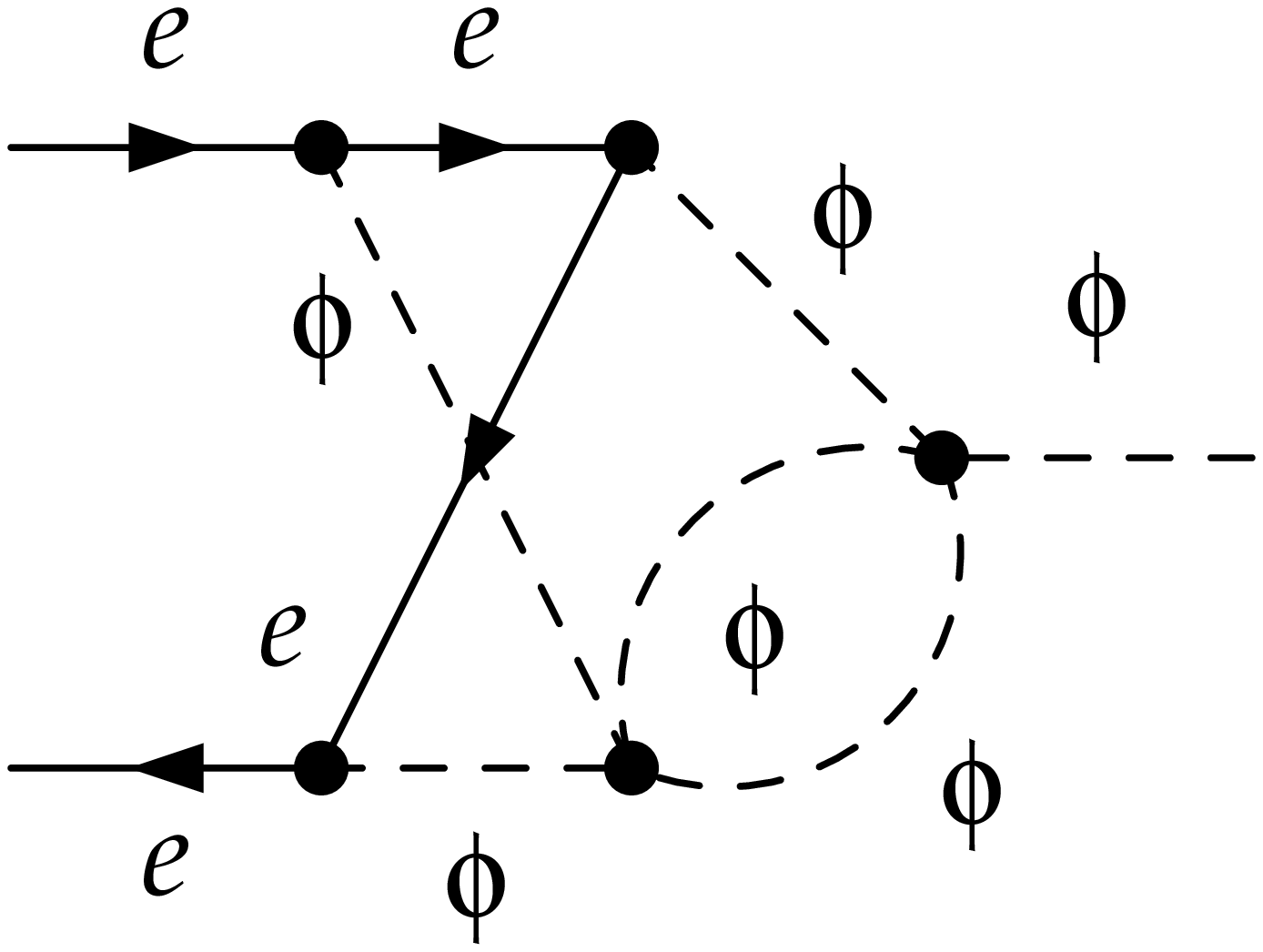}} & \vcenteredhbox{\includegraphics[scale=0.22]{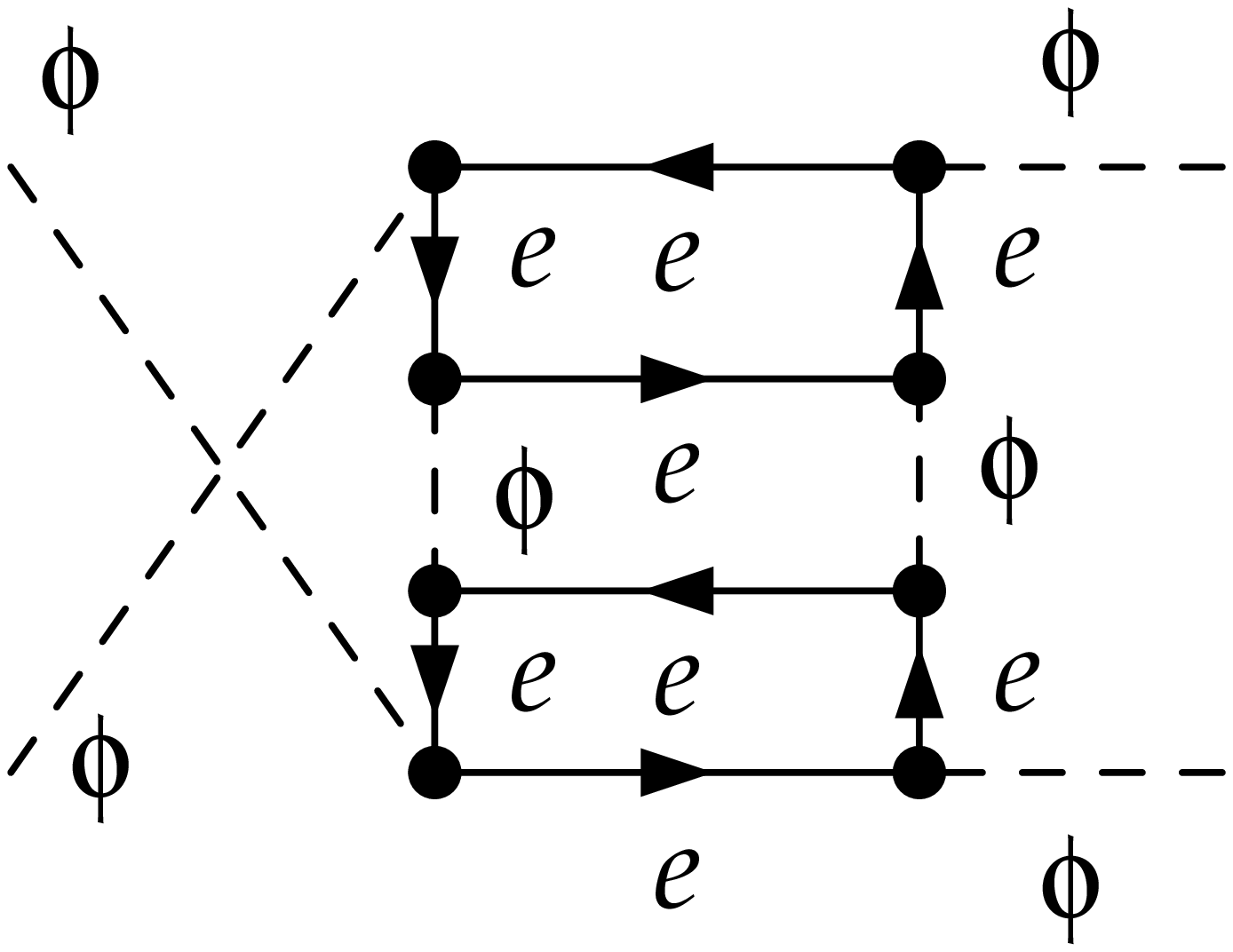}}
\end{tabular}
\caption{ Explicit diagram examples for the chiral Ising/Heisenberg Model\label{FIG:nPgreenHI}}
\end{figure}

In order to determine the $Z$-factors for the given model one  has to extract the UV-divergent pieces of the $n$-point Green function depicted in Fig.~\ref{FIG:nPgreen} for any suitable kinematics order by order in the loop expansion.
In Fig.~\ref{FIG:nPgreenHI} we show specific example diagrams for the chiral Ising and chiral Heisenberg case.

It turns out that one can factorize all diagrams appearing in the chiral Heisenberg case into a chiral Ising diagram times a $SU(2)$ spin weight factor when one uses projectors on the relevant $SU(2)$ structures.
The latter can therefore be calculated separately for each diagram and the transition from Ising to Heisenberg amplitudes can be done by multiplying diagram specific spin weight factor to the Ising diagram result.

Therefore the number of diagrams encountered in the Ising and Heisenberg case agree when considering Dirac fermions.
In Tab.~\ref{TAB:nDias1} we list the number of 1-PI Diagrams in dependence of the specific model, the number of loops $L$ and type of $n$-point function under consideration.
From the numbers it becomes clear that a hand calculation beyond two loops is not feasible anymore and one has to apply automated computer algebra to perform the calculation. 

\begin{table}
\small
\begin{minipage}{8cm}
\begin{tabular}{c|c|cccc|c}
$\lambda$ &$L$                     & $1$   & $2$    & $3$      & $4$      & $\Sigma$\\
 \hline
   \multirow{4}{*}{\vcenteredhbox{\rotatebox{90}{$\chi_{XY}$}}}&$ Z_{\phi^2},Z_{\phi}$ &  $2$  &  $9$   &   $112$  &   $2198$  &  $2321$\\
  &$ Z_{\psi}$            &  $2$  &  $14 $ &   $200$  &   $4014$  &  $4230$\\
  &$Z_{\psi\psi\phi}$     &  $2$  &  $41$  &  $1002$  &  $28701$  & $29746$\\
  &$Z_{\phi^4}$           &  $9$  & $173$  &  $5029$  & $147023$  & $152234$\\
 \hline \hline
   \multirow{4}{*}{\vcenteredhbox{\rotatebox{90}{$\chi_I$\&$\chi_H$}}}&$ Z_{\phi^2},Z_{\phi}$ &  $2$  &  $6$   &    $36$  &   $358$  &  $402$\\
  &$ Z_{\psi}$            &  $1$  &  $4$   &    $31$  &   $323$  &  $359$\\
  &$Z_{\psi\psi\phi}$     &  $1$  &  $11$  &   $145$  &  $2199$  & $2356$\\
  &$Z_{\phi^4}$           &  $9$  &  $93$  &  $1476$  & $26976$  & $28554$\\
\end{tabular}
\end{minipage}\begin{minipage}{4cm}
\begin{tabular}{c|c|cccc|c}
$\lambda$ &$L$                     & $1$   & $2$    & $3$      & $4$      & $\Sigma$\\
 \hline
   \multirow{4}{*}{\vcenteredhbox{\rotatebox{90}{$\chi_{XY}^*$}}}&$ Z_{\phi^2},Z_{\phi}$ &   $2$  &  $4$  &   $22$  &  $148$  & $176$\\
  &$ Z_{\psi}$            & $1$  &  $3$   &   $16$ &  $116$  & $136$\\
  &$Z_{\psi\psi\phi}$     &  $0$  &  $2$   &   $25$  &  $296$  & $323$\\
  &$Z_{\phi^4}$           &  $4$  &  $35$  &  $369$  & $4388$  & $4796$\\
 \hline \hline
   \multirow{4}{*}{\vcenteredhbox{\rotatebox{90}{$\chi_I^M$}}}&$ Z_{\phi^2},Z_{\phi}$ &  $2$  &  $5$  &   $27$  &   $213$  &  $247$\\
  &$ Z_{\psi}$            &  $1$  &  $4$   &   $29$  &   $283$  &  $317$\\
  &$Z_{\psi\psi\phi}$     & $1$  & $10$   &  $125$  &  $1779$  & $1915$\\
  &$Z_{\phi^4}$           & $6$  & $57$   &  $773$  & $12549$  &$13385$\\
\end{tabular}
\end{minipage}

\caption{ Number of diagrams encountered during $Z$-factor calculation in dependence of the model, loop number $L$ and specific $n$-point function.
Here the $\chi_{XY}^*$ model refers to the $\chi_{XY}$-Lagrangian which is defined in Ref. \cite{Zerf:2016fti} and $\chi_I^M$ is the chiral Ising Model assuming a Majorana instead of a Dirac fermion.
\label{TAB:nDias1}}
\end{table}

\section{Technical Details}
The calculation of the $Z$-factors is performed within a fully automated setup.
We chose the kinematics such that we have one small external momentum in case of the 2-point functions and
for 3- and 4-point functions all external momenta are set to zero.
To prevent the appearance of IR divergences the infrared rearrangement which was suggested in Ref.~\cite{Misiak:1994zw} and further developed in Ref.~\cite{Chetyrkin:1997fm} is used in order 
to rewrite all massless propagators in terms of massive propagators depending on a single artificial large mass via an exact decomposition which is only violated in finite pieces.
This is not a problem as long as the renormalization is done via explicit counter term insertion ensuring that all sub-divergences have been canceled before the overall divergence is determined.

Note that the same method has been used to calculate the five-loop QCD $\beta$-function and anomalous dimensions very recently~\cite{Luthe:2017ttg}\footnote{see also the corresponding proceedings contribution in here} and agrees with the result of a different evaluation method~\cite{Herzog:2017ohr,Chetyrkin:2017bjc}$^1$.
After rewriting all propagators one then is left with massive tadpole integrals in dependence of a single mass scale.
Up to including three loops one can use \textsc{MATAD}~\cite{Steinhauser:2000ry} to automatically reduce all of the integrals via integration-by-parts identities on the fly.\footnote{At the four loop level the program \textsc{FMFT}~\cite{Pikelner:2017tgv} has recently been published.}

The used setup runs through the following steps in order to arrive at an integrated result
\begin{enumerate}
\item  \textsc{QGRAF}\cite{NOGUEIRA1993279} is used to generate complete sets of Feynman diagrams.
\item \textsc{q2e} and \textsc{exp}\cite{Seidensticker:1999bb} are used to map all Feynman diagrams on single scale massive tadpole integral topologies and to generate \textsc{FORM} readable source files.
\item \textsc{FORM}\cite{Vermaseren:2000nd,Kuipers20131453,Ruijl:2017dtg} creates all counterterm insertions, performs the traces over the Clifford algebra, calculates the $SU(2)$ spin weight factors with the package \textsc{COLOR}\cite{vanRitbergen:1998pn}
and rewrites the amplitudes in terms of massive tadpole integrals with different powers of propagators.
Finally it reduces all appearing integrals to a set of master integrals with a predefined reduction table. 
At the four loop level there are nineteen master integrals\cite{Czakon:2004bu}.
\item The reduction table is created by \textsc{Crusher}\cite{crusher} up to including four loops and relies on integration-by-parts identities relating integrals with different propagator powers through a system of coupled equations to each other. 
This system of equations can be solved with the Laporta algorithm \cite{Laporta:2001dd} such that all appearing integrals can be written in terms of a linear combination of a finite number of master integrals.
\item A fully automated renormalization program written in \textsc{FORM} is used to extract the $Z$-factors order by order in the loop expansion from the bare amplitude results.
\end{enumerate}


\section{$\beta$- and $\gamma$-Functions }
The $\beta$-functions and anomalous dimensions of the fields can be directly obtained from the $Z$-factors.
After introducing the redefined coupling constants $g^2/(8\pi^2) \rightarrow y$ and $\lambda/(8\pi^2) \rightarrow \lambda$ they are given by:
\begin{align}
 \beta_\alpha=\frac{d\,\alpha}{d\ln \mu}\quad \alpha \in\{y,\lambda\}\,,\quad \gamma_x=\frac{d\ln Z_x}{d\ln \mu}\quad x\in\{\psi,\phi,\phi^2\}\,.
\end{align}
In order to keep track of the model $\lambda\in\{\chi_I,\chi_H,\chi_{XY}\}$ and the contribution from the $n$-th loop order up to four loops we write
\begin{align}
 \beta_{\alpha, \lambda}=&-\epsilon \alpha+\beta_{\alpha, \lambda}^{\text{(1L)}}+\beta_{\alpha, \lambda}^{\text{(2L)}}+\beta_{\alpha, \lambda}^{\text{(3L)}}+\beta_{\alpha, \lambda}^{\text{(4L)}}\,,\\
 \gamma_{x, \lambda}&=\gamma_{x, \lambda}^{\text{(1L)}}+\gamma_{x, \lambda}^{\text{(2L)}}+\gamma_{x, \lambda}^{\text{(3L)}}+\gamma_{x, \lambda}^{\text{(4L)}}\,.
\end{align}

As an example for the obtained results we show the $\beta$s and $\gamma$s for the chiral Ising model up to including 2 loops:
 \begin{align}
	\beta_{y, \chi\text{I}}^{\text{(1L)}}&=(3+2N)y^2\,,&
	\beta_{y, \chi\text{I}}^{\text{(2L)}}&=24y\lambda(\lambda- y)-\big(\frac{9}{8}+6N\big)y^3\,,\\
	\beta_{\lambda, \chi\text{I}}^{\text{(1L)}}&=36 \lambda ^2+4N y \lambda -N y^2\,,&
	\beta_{\lambda, \chi\text{I}}^{\text{(2L)}}&=4 N y^3+7N y^2 \lambda-72N y \lambda ^2-816 \lambda ^3\,,\\
	\gamma_{\psi, \chi\text{I}}^{\text{(1L)}}&=\frac{y}{2}\,,&
	\gamma_{\psi, \chi\text{I}}^{\text{(2L)}}&=-\frac{y^2}{16} (12 N+1)\,,\\
	\gamma_{\phi, \chi\text{I}}^{\text{(1L)}}&=2Ny\,,&
	\gamma_{\phi, \chi\text{I}}^{\text{(2L)}}&=24 \lambda ^2-\frac{5 N y^2}{2}\,,\\
	\gamma_{\phi^2, \chi\text{I}}^{\text{(1L)}}&=-12\lambda\,,&
	\gamma_{\phi^2, \chi\text{I}}^{\text{(2L)}}&=144 \lambda ^2-2 N y (y-12 \lambda )\,.
\end{align}
The shown two loop results are in agreement with the one obtained in Ref.~\cite{Rosenstein:1993zf}.
The recently evaluated three loop corrections for the chiral Ising and chiral Heisenberg model~\cite{Mihaila:2017ble}-- which had been obtained with \textsc{MATAD} -- were reproduced
with the self written table based Integrator.

For the chiral XY model the Feynman rules were implemented according to the stated Lagrange density in Eq.~(\ref{EQ:LXY}),
which yields a quite large number of Feynman diagrams (see Tab.~\ref{TAB:nDias1}) per loop level, because the fermion flow inside a closed fermion loop ($\sim N$) is not restricted.
That means for each diagram with a closed fermion loop there is always another diagram which agrees with the original one up to the reversed fermion flow within this very loop.
A more economic parametrization of the problem was used in Ref.~\cite{Zerf:2016fti} where the Lagrange density of the XY model has been written in terms of a double charged scalar cooper pair field
which involves the treatment of an indefinite fermion flow in each diagram such that two electrons can couple to a Cooper pair field.
The heavily reduced number of diagrams in this case can be found in Tab.~\ref{TAB:nDias1} for $\lambda=\chi_{XY}^*$.
However, both implementation yield the same $Z$-factors up to including four loops.
This is an expected result and serves as strong check, because both models are within the same universality class.
That means the three loop results presented here for the XY model are in agreement with the ones given in Ref.~\cite{Zerf:2016fti} which were obtained with full $N$ dependence but only published for $N=1/2$.

We refrain from explicitly showing all results up to including four loops in this contribution,
but refer the reader to Ref.~\cite{Zerf:2017zqi}.

\section{Critical Exponents at the QCP}
An IR fixed point exists for $\mu\rightarrow 0$ when the $\beta$-functions vanish for certain coupling values $\alpha^*=\{y^{*},\lambda^{*}\}$.
When the corresponding Jacobi matrix $[\mathcal{M}]_{ij}=\partial \beta_{\alpha_i}/\partial{\alpha_j}$ (which in this case is called stability matrix) has only positive eigenvalues $\omega,\omega^{'}$ ($\omega\leq\omega^{'}$) evaluated at $\alpha^*$,
then the corresponding fixed point is called stable.
In this case the Renormalization Group Equations (RGEs) make the couplings flow into a single fixed point from all directions in its vicinity with decreasing $\mu$.
If one or all eigenvalues is/are equal or smaller than zero the corresponding fixed point is called IR unstable and the RGEs will make the couplings flow away from it when $\mu$ is lowered.
The eigenvalues $\omega$ and $\omega^{\prime}$ are called stability exponent, where $\omega^{\prime}$ is less relevant, because it is larger than $\omega$.

We can now solve the $\beta$-functions order by order in $\epsilon$ for their zeros and obtain an expansion of $y^{*}$ and $\lambda^{*}$ in terms of polynomials in $\epsilon$.
Besides the trivial Gaussian $\alpha^*=\{0,0\}$ and Wilson-Fischer fixed point $\alpha^*=\{0,\lambda^{*}\}$ (which are unstable) there is (for all models) a non-trivial fixed point with $y^*\neq0, \lambda^*\neq0$.
From the result of the $\beta$- and $\gamma$-functions up to four loops one can determine the corresponding polynomial for $y^{*}$ and $\lambda^*$ in $\epsilon$ up to including the $\epsilon^4$ term.

This allows for the determination of anomalous dimensions at such a fix point $\gamma_x^*=\gamma_x(\alpha^*)$
and the correlation length exponent via
 \begin{align}
  \nu^{-1}=2+\gamma^{*}_{\phi^2}-\gamma_{\phi}^{*}\,.
 \end{align}
At the quantum phase transition the given exponents define the behavior of the corresponding order parameter $\xi$ in dependence of a reduced parameter $t$ to be
\begin{align}
	\xi \sim |t|^{-\nu}(1+C|t|^\omega+...)\,.\notag
\end{align}
The latter $t=(m_c-m)/m_c$ describes the deviation from a critical mass value $m_c$ at which the phase transition takes place.

As an explicit example for the obtained result one can set $N=1/2$ in the chiral XY model and investigate the fixed point at $y^*\neq0, \lambda^*\neq0$:
\begin{align}
	\gamma_\phi^* &=\gamma_\psi^* =\epsilon/3+\mathcal{O}(\epsilon^5)\,,\\
	\nu^{-1}&= 2-\epsilon+\frac{\epsilon ^2}{3}-\left(\frac{2 \zeta _3}{3}+\frac{1}{18}\right) \epsilon ^3+\frac{1}{540}\left(420 \zeta_3+1200 \zeta_5-3 \pi ^4+35\right) \epsilon ^4+\mathcal{O}(\epsilon^5)\,,\\
	\omega&=\epsilon-\frac{\epsilon ^2}{3}+\left(\frac{2 \zeta_3}{3}+\frac{1}{18}\right) \epsilon ^3-\frac{1}{540}\left(420 \zeta_3+1200 \zeta_5-3 \pi ^4+35\right) \epsilon ^4+\mathcal{O}(\epsilon^5)\,.\label{EQ:omegaXY}
\end{align}
That means the polynomials for the anomalous dimensions at the fixed point are the same for the fermion and the boson.
Further, there appear no contributions beyond the one loop level.
We further see $\nu^{-1}=2-\omega$.
It turns out that already the two $\beta$- and $\gamma$-functions for couplings and the fields do agree up to including four loops when setting $y$ equal to $\lambda$ (after a suitable redefinition see Ref.~\cite{thomas2005}).
One can further check that the $\beta$-function in this limit do reproduce the four loop Wess-Zumino result obtained in Ref.~\cite{Avdeev:1982jx}.
The reason for this is that at the IR fixed point the system is described by an emergent $\mathcal{N}=2$ super symmetric theory (SUSY). 
That means the fermion and scalar live in the same supermultiplet $\Phi$ and the $\mathcal{N}=2$ super symmetric Lagrangian involves only a single coupling.
The simple relation between $\omega$ and $1/\nu$ is in fact a consequence of the analytic property of the super potential in this Lagrangian.
Further, the SUSY property allows one to extract the correlation length exponent $1/\nu$ which depends by definition on the renormalization of the mass $m$ of the scalar, 
without calculating the relevant $Z_{\phi^2}$ (see Ref.~\cite{Zerf:2016fti}).
However, in the presented calculation $Z_{\phi^2}$ was explicitly calculated and provides a strong check of the result.

Concerning the stability of the fixed point one can see that the stability coefficient $\omega$ in Eq.~(\ref{EQ:omegaXY}) becomes negative when naively setting $\epsilon=1$ in order to extrapolate to $D=3$,
as soon as one includes the four loop term.
So in a naive/conservative approach to the stability question one cannot guarantee that the fixed point stays stable down to $D=3$ (like it would be the case when doing the analysis up to including three loops, only).
However, it is known that the obtained series can be of asymptotic nature and more sophisticated extrapolation are required.
As a first step one can employ the Pad\'e approximants $P_{[3/1]}$ or $P_{[2/2]}$ and indeed obtain a positive value for $\omega$ at $D=3$.
One can compare the results obtained with the two Pad\'es with the results from other methods in Tab.~\ref{FIG:critexp} and see that they fit quite well, considering the simplicity of the employed extrapolation.

\begin{table}
\small
\begin{minipage}[t][3.2cm][c]{7.5cm}
 \begin{tabular*}{\linewidth}{ c c c c c c}
    \hline\hline
$\chi_I(N=1/4)$  & & $1/\nu$ & $\eta_\phi$ & $\eta_\psi$ & $\omega$ \\
\hline
 $P_{[2/2]}$ & & 1.415 & 0.171 & 0.171 & 0.843\\
 $P_{[3/1]}$ & & 1.415 & 0.170 & 0.170 & 0.838\\[8pt]
FRG\cite{Gies:2017tod} ($R_1$) & &  1.385 & 0.174 & 0.174 & 0.765\\
FRG\cite{Gies:2017tod} ($R_2$) & &  1.395 & 0.167 & 0.167 & 0.782\\
CBS\cite{Iliesiu:2015akf}  & &  & 0.164 & 0.164 & \\
\hline\hline
\end{tabular*}
 \end{minipage}
\begin{minipage}[t][3.2cm][c]{7.0cm}
 \begin{tabular*}{\linewidth}{ c c c c c c}
\hline\hline
$\chi_{XY}(N=1/2)$  & & $1/\nu$ & $\eta_\phi$ & $\eta_\psi$ & $\omega$ \\
\hline
 $P_{[2/2]}$ & & 1.128 & 1/3 & 1/3 & 0.872\\
$P_{[3/1]}$ & & 1.130 & 1/3 & 1/3 & 0.870\\[8pt]
CBS\cite{Bobev:2015vsa}  & & 1.090 & 1/3 & 1/3 & 0.910\\
\hline\hline
\end{tabular*}
\vspace*{0.9cm}
\end{minipage}

\caption{Selected numerical results for the critical exponents at $D=3$ using regular Pad\'e approximants confronted with results from literature. 
FRG stands for Function Renormalization Group where $R_i$ corresponds to the usage of the $i$-th regulator.
CBS stands for Conformal Bootstrap.\label{FIG:critexp}}
\end{table}

In the chiral Ising model we run into a similar situation when setting $N=1/4$ (for fermion traces we use ${\rm tr} \mathds{1}_4 =4$).
Here the model runs at the non-trivial fixed point into an $\mathcal{N}=1$ super symmetric theory, where
in the original formulation (in three dimensions) the real scalar boson is the super partner of a two component Majorana fermion.
It turns out that this limit can naively be reproduced by choosing $N=1/4$ up to including three loops, only.
That means the SUSY induced relations like $\gamma_{\phi}^{*}=\gamma_{\psi}^{*}=\gamma^{*}$ and $2\nu^{-1}=D-\gamma^{*}$ hold only up to this order.

\begin{figure}
\centering
\includegraphics[scale=0.50]{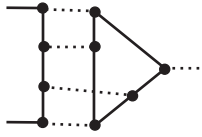}
\caption{ Example diagram in which DREG drops $\mathcal{N}=1$ SUSY relevant contributions $\sim \epsilon_{ijk} \epsilon_{lnm}$ \label{FIG:simepsilon2}.}
\end{figure}

Starting at four loops the $\beta$-functions become sensitive to the fact that we carry out our Clifford algebra around $D=4$ and not $D=3$ in which the $\mathcal{N}=1$ SUSY model is actually realized.
One could also say that the applied DREG breaks SUSY, because it automatically drops SUSY relevant contribution when setting all traces with an odd number of $\gamma_{\mu}$ to zero and
the SUSY relation between $\nu$ and $\gamma^*$ is not fulfilled anymore.
In more detail this happens in four loop Yukawa vertex correction diagrams like depicted in Fig.~\ref{FIG:simepsilon2}.
From a three dimensional realization of the Clifford algebra represented by the Pauli matrices $\sigma_i$ we know that we have a non-vanish contribution from traces with three $\gamma_{\mu}$ because ${\rm tr}(\sigma_i\sigma_j\sigma_k) \sim \epsilon_{ijk}$.
In the problematic diagrams the setup of the four loop momenta is sufficiently ``antisymmetric'' in order to retain a three dimensional $\epsilon$-tensor from each fermion chain.
Because any product of two $\epsilon$-tensors of rank three reduces to a fully anti-symmetric combination of three Kronecker $\delta$-tensors, we have a non-vanishing contribution to the amplitude which is set to zero in DREG.
Once we take this contribution of the relevant diagrams into account, for example through a calculation using an explicit $SU(2)$ clifford algebra reduction including the implementation of Majorana instead of Dirac fermions
(this slightly reduces the number of diagrams, see $\lambda=\chi_I^M$ in Tab.~\ref{TAB:nDias1}), we restore all SUSY relations at the four loop level.
Numerical results for the critical exponents at $D=3$ obtained through en extrapolation with Pad\'es can be compared to existing results from literature in Tab.~\ref{FIG:critexp}.

Beside the already mentioned non-trivial checks of the obtained result the agreement with the large $N$ limit calculations for GN models~\cite{Gracey:1990wi,Gracey:1992cp,Derkachov:1993uw,Vasiliev:1992wr,Vasiliev:1993pi,Gracey:1993kb,Gracey:2017fzu,Manashov:2017rrx} was ensured.
Further, the Yukawa coupling free terms reproduce the well known results of the $\phi^4$ theory at four loops, which are nowadays text book results  (see for e.g. \cite{Kleinert:2001ax}).
For a recent progress in $\phi^4$ theory at the six loop level see Ref.~\cite{Batkovich:2016jus}.
\vspace*{-0.3cm}

\section{Conclusions}
We have presented the perturbative determination of the $\beta$- and $\gamma$-functions for the chiral Ising, chiral XY and chiral Heisenberg GNY model at the four loop level in $D=4$ dimensions. 
In a first step we have employed Pad\'e approximants in order to obtain values for anomalous dimensions 
$\gamma^*_{\psi}$ and $\gamma^*_{\phi}$, the stability exponent $\omega$ and correlation length exponent $\nu$ for non-trivial IR fixed points in $D=3$ dimensions.
The obtained results are compatible with the existing predictions.
More sophisticated extrapolation methods systematically taking into account the asymptotic behavior of the obtained series
could help a lot to improve the extraction of values at $D=3$.
\vspace*{-0.3cm}

\acknowledgments
The author would like to thank the organizers for the well organized and nice conference at St.Gilgen,
M. Kraus for reading the manuscript and the collaborators I.~Herbut, B.~Ihrig, P.~Marquard, L.~N.~Mihaila, M.~Scherer for an uncomplicated and fruitful collaboration.
\enlargethispage{0cm}

\end{document}